# PAIRING INTERACTION AND TWO-NUCLEON TRANSFER REACTIONS

GREGORY POTEL

*CEA-Saclay, IRFU/Service de Physique Nucléaire, Gif-sur-Yvette, France*

ANDREA IDINI

*Institüt für Kernphysik, Technische Universität Darmstadt, Germany*

FRANCISCO BARRANCO

*Applied Physics Department III, University of Seville, Spain.*

ENRICO VIGEZZI

*INFN Milan, Italy*

RICARDO A. BROGLIA

*Department of Physics, University of Milan and INFN Milan, Italy*

*The Niels Bohr Institute, Copenhagen, Denmark*

*FoldLESs S.r.l., Monza, Italy*

## Introduction

Soon after the formulation of BCS theory [1,2], it was recognized by Bohr, Mottelson and Pines that the existence of an energy gap in the intrinsic excitation spectrum of deformed nuclei displayed a suggestive analogy with that observed in the electronic spectra of metallic superconductors and could, like this one, be described at profit in terms of correlated pairs [3]. Their paper represented the starting point of more than fifty years of experimental and theoretical BCS flavored studies of pairing in nuclei [4].

One of the important results which has emerged from this quest is that pairing has not one (bare nucleon-nucleon ($NN$) plus eventual $3N$ corrections cf. e.g. [5-8] and refs. therein) but two origins, the second one resulting from the exchange of collective nuclear vibrations between pairs of nucleons moving in time reversal states lying close to the Fermi energy (cf. [9-16] and refs. therein; see also [17]). This is why in discussing the pairing phenomenon one is simply forced to "complicate" the force through many-body correlations, a reflection of the retardation effects displayed by the nuclear pairing dielectric function.

In keeping with the fact that the building blocks of pairing correlations are Cooper pairs [18], two-nucleon transfer is specific to probe them, the associated absolute differential cross sections being the main, model independent observables relating theory with experiment.

In the first part of the present contribution we report on recent progress made within this context [19-22], progress which has allowed to shed light into the interplay of the bare $NN$ pairing interaction with collective vibrations and to obtain, *inter alia*, quantitative evidence of phonon mediated pairing in halo exotic nuclei (cf. [23-26] and refs. therein ). This is the subject of the second part of the paper.

## Pair transfer and pairing correlations in nuclei

At the basis of BCS theory of superconductivity one finds the condensation of strongly overlapping Cooper pairs, a model which has been applied with success to the description of pairing correlations in atomic nuclei. There is however a main difference, as compared with the case of the condensed matter scenario in which BCS theory originated. In the nuclear case, fluctuations of the pairing field as well as of the normal density are very important and renormalize in a conspicuous way the different quantities entering the theory. In particular, around closed shell nuclei, systematic evidence exists of the correlation and stability of the pair addition (cf. Fig. 1) and pair subtraction modes which are strongly excited in two-particle transfer reactions. Pairing vibrations (cf. [27-29] and refs. therein), the nuclear embodiment of single Cooper pairs, smooth out through zero-point fluctuations (ZPF) the sharp change of the occupancy of levels around the Fermi energy (Fig. 1a), bottom), taking place in mean field, thus paving the way for an eventual phase transition from normal to superfluid phases.

A number of pairing vibrational bands have been observed throughout the mass table, containing up to three phonon states [30,31]. Because of the strong correlations displayed by these vibrational modes, their microscopic properties can be accurately described in terms of the RPA and of a constant pairing strength, leading to reliable values of the X- and Y-amplitudes (cf. Fig. 1), and thus of two-nucleon spectroscopic amplitudes (cf. also ref. [31], Tables XVI-XVIII). The study of pairing vibrations provides, among other things, insight into the mechanism by which a nuclear superfluid phase eventually emerges from the condensation of pairing vibrational modes, as the system under study moves progressively away from closed shell nuclei. The condensation of these extended and thus strongly overlapping, bosonic objects gives rise to a highly correlated superfluid state, displaying overall phase coherence. Superfluidity is tantamount to the existence of a finite ground-state average value of the pair addition and removal operators $P^+$, $P$ in the ground state, that is, to a finite value of the order parameter

$$\alpha_0 = |\langle BCS|P^+|BCS\rangle = |\langle BCS|P|BCS\rangle$$

a quantity which provides an estimate of the number of correlated pairs in the BCS ground state ($\approx$ 4-8).

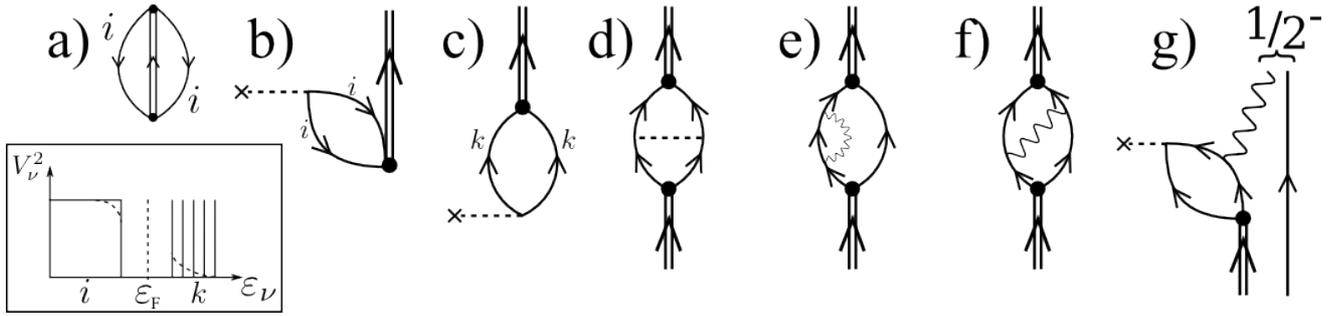

*Figure 1*. RPA, Nuclear Field Theory (NFT) diagramatic representation of the structure [54,55] and reactions [39] of, and with pair addition modes. This pairing vibration is mainly a correlated superposition of two-particle states with (cf. c)) forwardsgoing amplitudes $X_k$ on the different orbitals above the Fermi energy ( $\varepsilon_k > \varepsilon_F$ ) . The possibility of creating this state by populating hole states below the Fermi energy (cf. b)), with backwardsgoing amplitudes $Y_i$( $\varepsilon_i \leq \varepsilon_F$ ) arises from the presence of two-particle, two-hole configurations in the ground state of the closed shell system (cf. a), top)), ZPF which smooth out the discontinuity in level occupancy at $\varepsilon_F$ (cf. a),bottom). The solid dot represents the strength and form factor with which particles, moving in time reversed states, couple to the collective, quasi-boson pairing degree of freedom. It results from the combined effect of a four-point vertex (bare interaction), see graph d), and of vertex correction (induced interaction) processes, an example of which is given in diagram f). Diagram e) is representative of processes which dress the single-particle states. By intervening processes e) and f) with an external field which picks up two nucleons from the system, one can force the virtual phonon to become a real final state. Assuming that the pair addition mode is the two-neutron halo of $^{11}$Li, the quadrupole vibration of the $^{8}$He core, coupled to a $p_{3/2}$ ($\pi$) proton state, the process g) describes the population of the first excited state of $^{9}$Li in the reaction $^{1}$H( $^{11}$Li, $^{9}$Li (1/2-;2.69 MeV)) $^{3}$H [23,24].

It also gives a measure of deformation in gauge space, the counterpart of deformation in ordinary space (cf. e.g. [17,28,32] and refs. therein). Just as adjacent (I,I±2) states lying along a, e.g., quadrupole rotational band are connected by strongly enhanced values of the quadrupole operator, the adjacent $0^+$ ground states (N,N±2) of e.g. a chain of superfluid isotopes are connected by strongly enhanced values of the pair transfer operator, measured in terms of single- and two-particle units respectively [28,33]. This result testifies to the fact that these $0^+$ states are members of a pairing rotational band (see e.g. Fig. 2). Within this scenario, pairing vibrations and rotations together with single-particle motion and vibrations and rotations in "normal" (three-dimensional) space constitute elementary modes of excitation.

The suggestive analogy concerning the nuclear phenomena associated with spontaneous symmetry breaking in 3-D and in gauge space (cf. e.g. Table XI, ref. [31]) although extending also to the reaction (decay) processes in which these rotational modes are specifically probed, is not operative as far as the calculational details are concerned. In fact, Coulomb excitation (electromagnetic decay) and Cooper pair transfer display very different levels of calculational challenges (complexity) concerning their implementation. This is keeping with the fact that in Coulomb excitation, let alone electromagnetic decay, one has to deal with a single mass partition, a fact which makes it possible to treat structure and reaction, to a large extent, separately. This is not the case for two-nucleon transfer reaction, in which case mass partition is different between entrance and exit channels, a fact which leads to recoil effects and thus to an important coupling between relative motion (reaction) and intrinsic motion (structure). In fact, the situation is even richer, in keeping with the fact that nucleons may be transferred not only simultaneously but also successively. Thus, one is confronted in the calculation of the absolute value of two-nucleon transfer cross sections, with the opening of a new channel and thus of a new mass partition (e.g. (N+2) + p $\rightarrow$ (N+1) + d $\rightarrow$(N) + t ). It is then not surprising that the theory of Coulomb excitation and electromagnetic decay was quantitatively operative only few years after the first observation of rotational bands [28,34], while it took decades (cf. Fig.10 of ref. [19] for an overview of the groups and the practitioners involved in the quest) after the first observation of a pairing rotational band [35-38] before one was able to calculate absolute Cooper pair transfer cross sections which account for the observations within experimental errors (see Fig. 3).

The fact that, as a rule, successive transfer dominates over simultaneous transfer, and that, in both processes, the transferred nucleons display equivalent pairing correlations, is a consequence of the fact that Cooper pairs are weakly bound ($<<\varepsilon_F$), highly extended ($>>R$) objects. Consequently, the minimum theory of two-nucleon transfer corresponds to second order DWBA, where the two above mentioned processes are taken into account properly corrected by non-orthogonality effects (cf. e.g. [39], cf. also [19] and refs. therein, in particular those associated with Fig. 10 of this reference). It is only recently that these well known elements were implemented into a versatile software [40] with which, making use of well tested, state of the art spectroscopic amplitudes, and global optical potentials, one can calculate absolute two-particle transfer differential cross sections which account for the experimental findings within experimental errors throughout the mass table [19-22, 24,25]. Examples of these quantitative results are displayed in Fig. 3 (cf. also Figs. 5 and 6 as well as Table 3 of ref. [19]).

It is of notice the essential difference existing in the physics which is at the basis of this agreement concerning the two theoretical results shown in the upper and middle panels ("upper") of Fig.3, as compared to the two lowest panels, in particular the lowest right panel ("lowest"). In fact, the results displayed in "upper" depend little on the details of the pairing interaction employed, or the exact value of the energies and $Z$-values (see below) of the single-particle levels used in the calculations, a fact intimately connected with the constancy of the lowest quadrupole mode through the Sn-isotopes (evidence of the validity of generalized seniority), and of the large two-neutron separation energy associated with $^{208}$Pb. This is the reason why simple models like BCS or RPA which embody the physics of coherent pairing modes, i.e. pairing rotations and vibrations, provide essentially "exact" two-nucleon spectroscopic amplitudes (i.e. $U_\nu V_\nu$ and $(X_j, Y_j)$ factors). On the other hand, the results displayed in "lower" are very sensitive to the details of the single-particle energies and associated $Z$-values, as well as to components in the $^{11}$Li ground state wavefunction displaying a 1% probability, a result of rather refined NFT calculations.

In keeping with the above parlance, the type of results displayed in "upper" and "lower" provide confidence in the fact that one now knows how to accurately calculated absolute two-nucleon differential cross sections in (nuclear structure) simple cases as well as to quantitatively predict new mechanisms to dynamically violate gauge invariance.

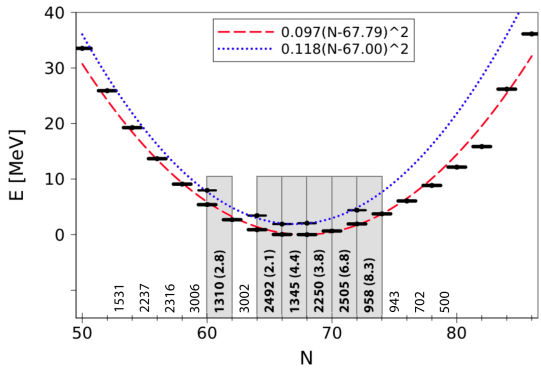

**Figure 2.** *Experimental energies (plus a linear N-dependent term introduced for convenience) of the $0^+$ states of the Sn isotopes (ground states and pairing vibrations). The dashed and dotted lines represent the parabolas given in the insets, corresponding to the ground state and to the (average) excited state-based pairing rotational bands. The absolute values of the (gs)→ (gs) integrated cross sections (in µb) are given (perpendicular) to the abscissa, as a function of N. In the shaded areas we report the experimental values [41-45], while the remaining values correspond to theoretical predictions integrated in the range $0° < \Theta_{CM} < 80°$. (cf. also Fig.3). For the first group (experimental) we also report the relative (p,t) pairing vibrational cross sections (in %), normalized with respect to the ground state cross sections (after ref. [20]).*

Let us now elaborate on the fact that the coherent character of pairing correlations manifests itself equally well in simultaneous than in successive transfer processes. In fact, for superfluid nuclei the quantity (1) is given, in the case of simultaneous transfer, by the relation

$$\alpha_0 = \sum_{\nu>0} \langle BCS | a_\nu^+ a_{\bar\nu}^+ | BCS \rangle \text{ and by the expression}$$

$$\alpha_0 = \sum_{\nu,\nu'>0} \langle BCS | a_\nu^+ | int(\nu') \rangle \langle int(\nu') | a_{\bar\nu}^+ | BCS \rangle$$

in the case of successive transfer. Inserting $|int(\nu)> \approx \alpha_\nu^+ |BCS>$ and making use of the quasiparticle transformations both relations lead to $\sum_{\nu>0} U_\nu V_\nu$. This result is intimately connected with the large distance (correlation length $\xi \approx \hbar\, v_F/\Delta \approx 36$ fm) over which Cooper pair partners correlate.

**The pairing interaction and medium polarization effects**

The nature of the attractive pairing force acting between electrons represented a central question in the development of the theory of superconductivity in metals (screened Coulomb field plus electron-phonon mediated interaction, cf. [46] and refs therein). In the nuclear case the bare *NN*-interaction is strongly attractive in the $^1S_0$ channel, and mean field calculations lead to neutron or proton pairing gaps of the order of those derived from experimental data. It is of notice, however, the latest developments concerning *3N* (mainly repulsive) corrections to the bare *NN*-interaction [5-6].

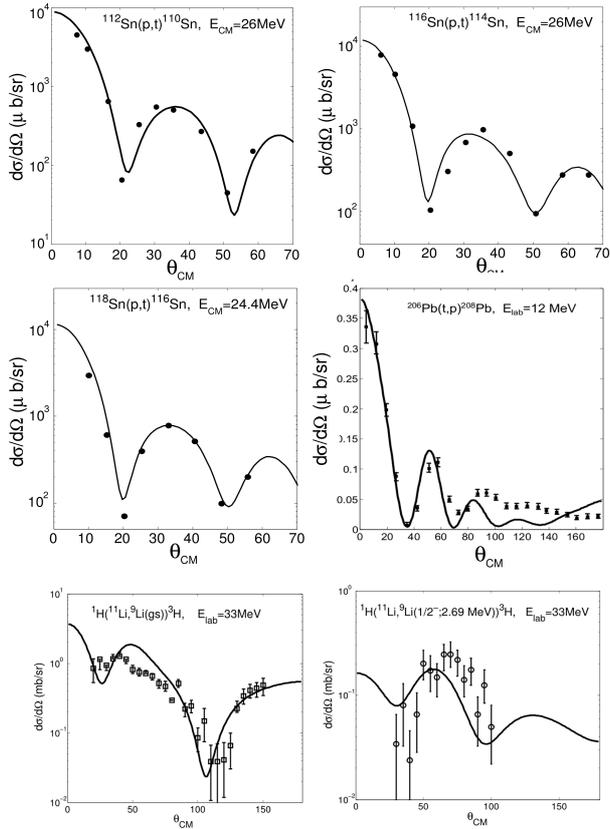

**Figure 3**. *Absolute cross sections associated with two–neutron transfer reactions involving superfluid (Sn-isotopes) and pair vibrational ($^{208}$Pb and $^9$Li) nuclei. Making use of the "exact" two-nucleon spectroscopic amplitudes (cf. text), of global optical potentials, and of two–nucleon transfer software developed within the framework of second–order DWBA [40], the corresponding absolute differential cross sections associated with these reactions were calculated and are displayed (continuous curves) in comparison with the experimental data [19,20,23,24,29,41-43]. It is of notice that the absolute differential cross section associated with the $^9$Li(1/2$^-$, 2.69 MeV) state provides a realization of the NFT process depicted in Fig. 1g) in direct comparison with the data (right bottom panel).*

Before dealing with the question of how two-nucleon transfer reactions can shed light on the question of the interplay (relative importance) of bare and induced pairing interaction, let us remind the basics of medium polarization effects. The relevance of these effects in connection with one-nucleon transfer reactions have been recognized since a long time (cf .e.g. [47] and refs. therein). In such reactions, as well as in (*e,e'p*) processes, one often observes that the single-particle strength associated with levels lying close to the Fermi energy are fragmented over a number of peaks, and the single-particle content of the main peak varies typically from 60% to 80% of the value expected in the independent particle limit.

Part of the reduction of single-particle strength can be ascribed to the short-range part of the *NN*−interaction (short wavelength mechanism) which shifts single-particle strength away from the Fermi energy (high momentum processes). Another part of the reduction is associated with a long wavelength mechanism resulting from the interweaving of single-particle and low-lying collective vibrations (low-*k* processes). Examples of such processes are displayed in Fig. 4. A nucleon can bounce inelastically off the nuclear surface, setting it into vibration, changing its state of motion and, at a later time, by reabsorbing the vibration returns to its original state as shown in (a). Important effects are also connected with the process depicted in (b), obtained from a time ordering from process (a). It leads to a partial blocking of the ground state correlations (oyster-like diagram), process giving rise to an effect known in atomic physics as the Lamb shift [48].

Through the processes displayed in Fig. 4, a nucleon moving in a single-particle configuration, is forced into more complicated configurations. In other words, the single-particle strength becomes fragmented, and the discontinuity of the occupation numbers at the Fermi energy, *Z=1* in the case of the non-interacting system, is reduced (*Z<1*) .

The probability with which the associated components of the ground state wavefunction containing phonon degrees of freedom are present in the dressed single-particle states can, in principle, be experimentally determined in one-particle transfer processes populating the excited states of the A-1 system. As an example, we refer to the $p(^{11}Be, ^{10}Be(2^+))d$ reaction [49]. The presence of such components has been shown to be relevant also in break-up reactions [50]. We note that the importance of contributions of multi-step processes, which can populate the final states in question, even in absence of correlations in the initial state, must be quantitatively assessed. In fact, the possibility of observing the excitation of states associated with the "complex" components of the single-particle wavefunction of the initial ground state, is connected with situations in which multistep processes are hindered by structure and/or *Q*−value effects.

Renormalization effects of the nuclear pairing gaps have been discussed for quite some time in connection with infinite matter (cf. [51] and refs. therein). Work started at the end of the 90's provided evidence through the result of detailed calculations that in finite nuclei, the exchange of virtual phonons - in particular quadrupole and octupole surface vibrations - between two neutrons coupled to $J^\pi = 0^+$ gives rise to an energy dependent attractive force leading to state dependent pairing gaps which, in average account for a conspicuous fraction of the empirical values [9-17] (within this context cf. also ref. [28], p.432). The processes at the basis of the induced pairing interaction are depicted in Fig. 4(c): a vibration excited by a nucleon is reabsorbed by a second nucleon. Such a process leads to an induced interaction among nucleons, associated with the polarization of the nuclear medium.

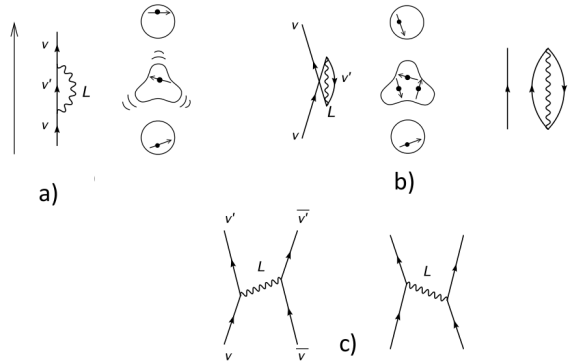

**Figure 4**.

*NFT diagrams describing the lowest order processes which renormalize both normal and abnormal densities: (a) polarization, (b) correlation, (c) induced pairing interaction, processes. For a concrete embodiment of these process and their relation to experiment cf. Fig. 1g), Fig. 3 (bottom) and Figs. 6(I),6(II).*

The superposition of the bare and the phonon induced interactions ($V^{eff} = V^{bare} + V^{ind}$) tends to increase the value of the Cooper pair binding energy as compared to the $V^{bare}$ result, but the coupling of surface phonons to single-particle states leads to a depopulation of the pure single-particle states through self-energy processes (cf. Figs. 4a) and b)). As a consequence, the BCS gap equation is modified by the presence of $Z$-factors, a well known effect in the theory of superconductivity [46], leading to [13,15-17]:

$$\widetilde{\Delta}_1 = -Z_1 \sum_2 \frac{V^{eff}(12) Z_2 \widetilde{\Delta}_2}{2 E_2}.$$

One can then identify two contributions to the gap, $\widetilde{\Delta} = \widetilde{\Delta}^{bare} + \widetilde{\Delta}^{ind}$. The effects of the basic renormalization diagrams can be taken into account up to infinite order, by solving the Nambu-Gor'kov equations, leading to a consistent theoretical picture, that accounts for these effects both on the single-particle motion and on the the pairing interaction. An example of such calculations for the ground-state pairing gap is shown in Fig. 5. The total gap $\widetilde{\Delta}$ is considerably larger than the value $\Delta^{BCS}$ obtained solving the usual BCS equations with the bare (Argonne) pairing force without taking into account renormalization of single-particle motion.

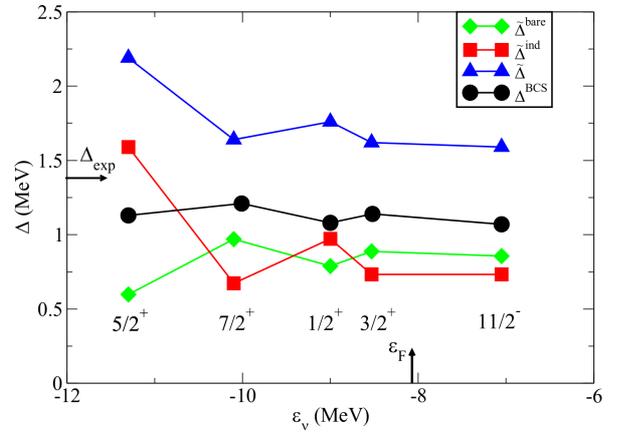

**Figure 5**. *State-dependent pairing gap in $^{120}$Sn calculated by solving the Nambu-Gor'kov equations. Single-particle levels were obtained from a Hartree-Fock calculation with the Skyrme SLy4 interaction. The Argonne interaction was used as the bare pairing force.*

The contributions coming from the bare and from the induced interactions to $\widetilde{\Delta}$ are comparable. At the Fermi energy the value of $\widetilde{\Delta}$ is larger than experiment by ≈ 20%. It is of notice that coupling to spin modes will somewhat reduce the value of $\widetilde{\Delta}$. However, at present no complete microscopic calculation of the pairing gap including both the bare interaction and medium polarization effects exists. Main open problems remain the determination of the initial mean field, the role of three-body forces and the coupling to spin modes.

Theory indicates that the induced interaction is concentrated around the Fermi energy and is strongly surface peaked. It is however not straightforward to have direct information of these properties: its effects can, in many cases, be simulated by adjusting the strength of the bare interaction. In fact, the spatial dependence of the Cooper pair, at least in well bound nuclei, depends only weakly on the details of the pairing interaction. Within this context one can posit that the pairing gap, although intimately connected with pairing in nuclei, is not the specific quantity to probe the corresponding correlations, at least as far as the nature of the interaction that generates them is concerned. This is also in keeping with the fact that the pairing gap is a derived quantity (e.g. 3-point empirical value, requiring the knowledge of three different nuclear masses). On the other hand, with the help of two-nucleon transfer reactions, one can force the virtual processes displayed in Figs. 1e),f) to become final, observable states. In fact, being able to accurately calculate absolute differential cross sections, information about the phonon admixture in the Cooper pair wavefunction can be obtained by studying pair transfer to excited collective vibrational states of the core (Fig. 1g) .

Let us conclude this section with a technical note. At variance with infinite systems in general, and condensed matter in particular, in which case particle number fluctuations are negligible, in the nuclear case they play an important role. This is the reason why much work has been dedicated to this question (projection methods, RPA tecniques etc., cf. [17,52,56,57] and refs. therein). Within this context, the pairing gap becomes

$\Delta = (\tilde{\Delta}^2 + G^2 \, S_0 \, (RPA)/2 \,)^{1/2}$, where $S_0(RPA)$ contains the (particle-conserving) matrix elements of $P^+$ and P (cf. e.g. ref. [17], p.151). While projection effects are dominant at the phase transition, they lead to corrections of the order of 10-20% for the ground state pairing gap.

**The case of halo nuclei**

Renormalization effects can have particularly striking consequences in halo nuclei like $^{11}$Li, systems which are weakly bound and easily polarizable. In particular, it was proposed [53] that the coupling of single-particle levels to quadrupole vibrations of these systems plays an important role to explain the positive parity of the ground state of $N = 7$ isotones, a dynamical effect going beyond mean field theory. The particle-vibration matrix elements associated with quadrupole vibrations are, in these nuclei, very large (see Fig. 6(I)). In fact, the neutron $2s_{1/2}$ orbital is shifted downwards by several MeV by virtue of its coupling to configurations of the type $[d_{5/2} \otimes 2^+]_{1/2^+}$ (polarization diagram (a) in Fig.

4). Furthermore, the neutron $1p_{1/2}$ orbital is shifted upwards as a result of the suppression of ground state correlations (Pauli principle processes) mostly associated with the configuration $[p_{1/2} \otimes p_{3/2}^{-1}]_{2^+} \otimes 2^+]_{0^+}$ (correlation diagram (b) in Fig. 4).

A dynamical Nuclear Field Theory ( NFT, cf. [54,55] and refs. therein) description of the two-neutron halo nuclei $^{12}$Be and $^{11}$Li, based on the coupling to the vibrations of these systems and of their cores, provides an overall account of their nuclear structure properties [26,58]. Dealing with a single, dressed Cooper pair, the corresponding wavefunction can be obtained by summing the processes shown in Figs. 1e),f) to infinite order with the help of Dyson's equation, a treatment of the variety of couplings tantamount to a full diagonalization, including the (discretized) continuum. In fact, and as is well known, the continuum plays an essential role in the case of $^{11}$Li, for which all the relevant single-particle orbitals are resonant or virtual states, in keeping with the fact that $^{10}$Li is unbound.

Furthermore in $^{11}$Li, an important role is played by the low-lying dipole state (pigmy resonance ≈1 MeV), responsible of much of the glue binding the neutron halo Cooper pair to the $^9$Li core. The resulting wavefunction of the dressed neutron halo can be written as shown in Fig. 6 (b) and (II).
It turns out that the (short range) bare $^1S_0$ neutron-neutron pairing interaction leads, in the present case, to a small contribution. This in keeping with the very low angular momenta available to the neutrons (essentially $s$, $p$ states being involved in the very extended and diffuse $^{11}$Li halo). The wavefunction of the $3/2^-$ ground state of $^{11}$Li is then obtained by coupling the $p_{3/2}(p)$ proton, treated as a spectator, to the neutron halo.
A detailed analysis of the reaction $^1H(^{11}Li,^9Li)^3He$ reaction performed at TRIUMF [23] with a $^{11}$Li beam (inverse kinematics) has been carried out [24]. Two states were observed: the $^9$Li ground state and the first $^9$Li($1/2^-$) excited state, which is interpreted as the lowest member of the $p_{3/2}(\pi) \otimes 2^+$ multiplet. It is of notice that the angular distribution associated with the ground state is very sensitive to the relative weight of the $s^2$ and $p^2$ configurations in the wavefunction displayed in Fig. 6b) and 6(II), wavefunction which reproduces quite accurately the experimental findings (cf. Fig. 3).

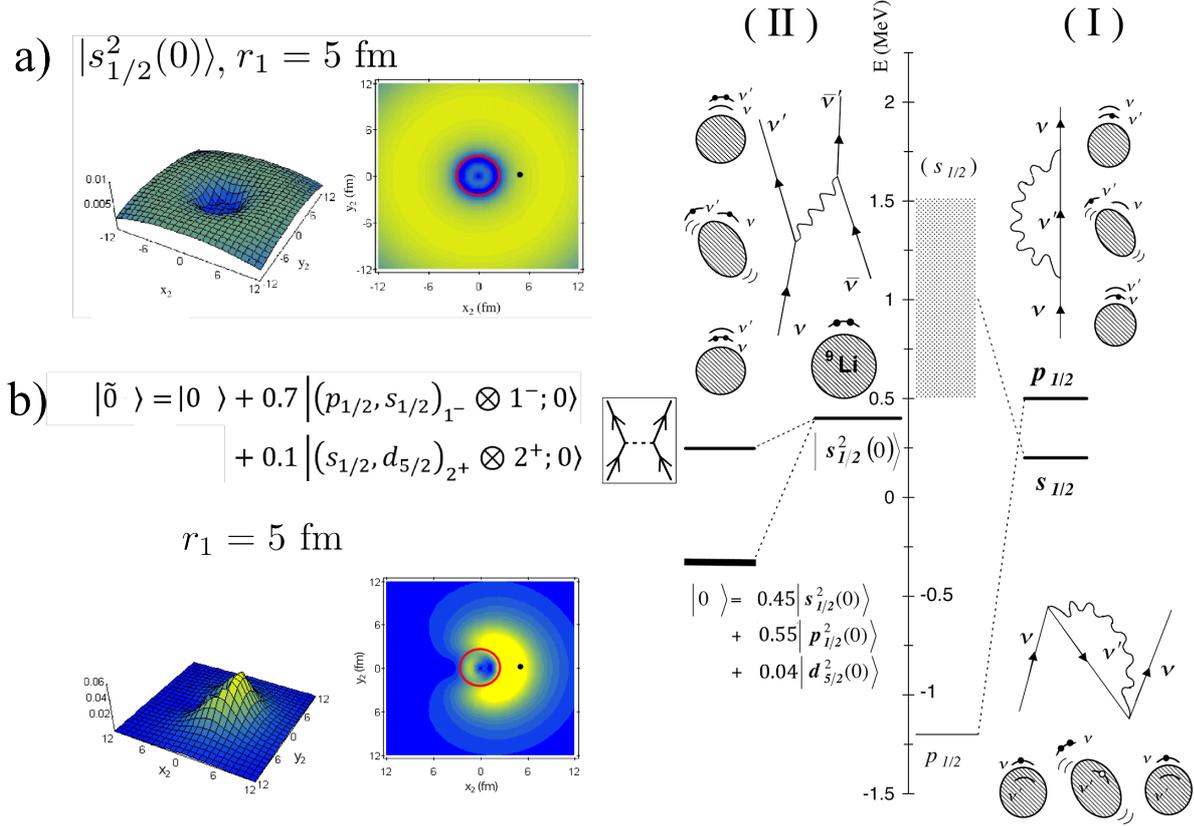

**Figure 6.** *The NFT scheme synthetized in Figs. 1a)-f) and Fig. 4a)-c) becomes operative concerning the structure of $^{10}$Li and $^{11}$Li: (I) self-energy processes, giving rise to parity inversion in $^{10}$Li; (II) bare (boxed inset) and induced pairing interaction binding the halo neutron pair to the $^{9}$Li core, through a bootstrap mechanism, in which the neutrons exchange the pigmy dipole resonance of $^{11}$Li, as well as the quadrupole vibration of the core, as testified by the wavefunction b).*

*In other words, the color snapshots displayed in a) and b) attempt at describing the becoming of the neutron halo Cooper pair of $^{11}$Li, from an uncorrelated $s_{1/2}^2(0)$ configuration to a strongly correlated, (weakly) bound two-neutron state. It is of notice that the bare interaction ( boxed inset in (II)), corresponding to the process depicted in Fig. 1d) (NFT four point vertex, rule (II) of NFT, cf. e.g. ref. [55], p. 314) lowers the $s_{1/2}^2(0)$ (as well as the $p_{1/2}^2(0)$ ) pure configurations by only 100 keV, and is not able, by itself, to bind the pair. The color plots display the modulus square of the two-neutron wavefunction as a function of the coordinates of the two nucleons (left) and the probability distribution of one neutron with respect to the second one held fixed on the x-axis (at a radius of 5fm, solid dot). The red circle schematically represents the core. After [58].*

The renormalization processes in which the neutrons of the halo Cooper pair of $^{11}$Li either emit and reabsorb a collective (p-h)-like quadrupole vibration (effective mass processes, Fig. 1e)) or exchange a phonon (vertex correction, Fig. 1f)) can, in a two-particle pick-up reaction (Fig. 1g)), populate the first excited state $1/2^-$, lowest energy ($E^* = 2.69$ MeV) member of the $|2^+ \otimes \pi(3/2^-)\rangle_{J^\pi,(J^\pi=1/2^-,...,7/2^-)}$ multiplet of $^9$Li. The absolute value of the corresponding two-nucleon transfer cross section provides an accurate measure of the probability with which the $|(s_{1/2},d_{5/2})_{2^+} \otimes 2^+; 0\rangle$ component appears in the $^{11}$Li ground state (cf. Fig 6b)) and thus of the role the quadrupole vibration plays in binding the neutron halo Cooper pair. This is also in keeping with the fact that alternative channels, like final-state inelastic excitation and neutron break-up, lead to negligible contributions [24].

The fact that theory reproduces the observed absolute differential cross sections, testifies to the fact that *NFT of* structure and reactions [39,54,55], is able to accurately predict [58] and describe [24] the consequences of the induced nuclear pairing interaction.

While this result can, arguably, be considered a milestone in the understanding of the origin of pairing in nuclei, we feel equally important and timely the developments taking place at a breathtaking pace, concerning the connection of the *NN*-bare interaction and of the quark degrees of freedom, and of its regularization in terms of renormalization group methods or similar techniques ($V_{low-k}$), to work out a pairing interaction (taking also *3N* terms into account), which can be used in nuclear structure calculations. It is likely that these developments will contribute together with the ones presented above, in an important and hopefully conclusive way to the quest of assessing the relative role of bare and medium polarization effects in the nuclear pairing interaction.

**Acknowledgment**

We thank Luisa Zetta and Paolo Guazzoni as well as Ritu Kanungo and Isao Tanihata for discussions and clarifications concerning their state of the art $^{A+2}$Sn(p,t)$^A$Sn and $^1$H($^{11}$Li, $^9$Li)$^3$H data, respectively. Collaboration with Ben Bayman is gratefully acknowledged. RAB acknowledges his debt towards Daniel R. Bès for many discussions and clarifications concerning the physics which is at the basis of the subjects treated in the present contribution.